\documentclass[pra,aps,twocolumn]{revtex4}
\usepackage{graphicx}

\begin{document}
\title{Reconfigurable photoinduced metamaterials in the microwave regime}

\author{Carlo Rizza$^{1,2}$, Alessandro Ciattoni$^2$, Francesco De Paulis$^3$, Elia Palange$^4$, Antonio Orlandi$^3$, Lorenzo Columbo$^5$ and Franco Prati$^1$}

\affiliation{$^1$Dipartimento di Scienza e Alta Tecnologia, Universit\`a dell'Insubria, via Valleggio 11, Como, I-22100 Italy}
\affiliation{$^2$Consiglio Nazionale delle Ricerche, CNR-SPIN, via Vetoio 10, L'Aquila, I-67100 Italy}
\affiliation{$^3$Dipartimento di Ingegneria industriale e dell'informazione e di Economia, via Giovanni Gronchi 18, L'Aquila, I-67100 Italy}
\affiliation{$^4$Dipartimento di Fisica e Chimica, via Vetoio 10, L'Aquila, I-67100 Italy}
\affiliation{$^5$Consiglio Nazionale delle Ricerche, CNR-IFN, via Amendola 173, Bari, I-70126 Italy}

\date{\today}

\begin{abstract}
We investigate optically reconfigurable dielectric metamaterials at gigahertz frequencies. More precisely, we study the microwave response of a subwavelength grating optically imprinted into a semiconductor slab. In the homogenized regime, we analytically evaluate the ordinary and extraordinary component of the effective permittivity tensor by taking into account the photo-carrier dynamics described by the ambipolar diffusion equation. We analyze the impact of semiconductor parameters on the gigahertz metamaterial response which turns out to be highly reconfigurable by varying the photogenerated grating and which can show a marked anisotropic behavior.
\end{abstract}

\maketitle

\section{Introduction}

Over the past few years the research interest on metamaterial science has shown
a sudden increase since metamaterials offer an enormous potential to engineer
at will the electromagnetic response. Metamaterials are composite materials
built by embedding subwavelength inclusions within a host medium or by
depositing them onto a suitable substrate \cite{Cai}. One of the first
metamaterial examples used to mold the light flux to yield a prescribed
behavior is Pendry's lens, a metamaterial device capable of producing
sub–diffraction-limited images \cite{Pendry_1,Fang}. In order to achieve
super-resolution, many authors have suggested structures supporting
diffractionless waves such as far-field optical hyperlenses \cite{Jacob,Liu},
multilayered metamaterials operating in canalization regime \cite{Belov} and
Kapitza metamaterials \cite{Rizza_1}. Metamaterials are characterized by their
unusual effective electromagnetic parameters as, for example, very large or
very small permittivity or permeability (e.g. epsilon-near-zero or
epsilon-infinity media) \cite{Elser,Silveirinha,Alu_1,Jin,Ciattoni_2}. In
addition, chiral \cite{Wang} and non-reciprocal metamaterials \cite{Popa}
showing strong spatial non-local response have been conceived. On the other
hand, exploiting the wide metamaterial flexibility, Pendry et al. proposed
transformation optics \cite{Pendry_2} as a new scheme to steer light at will.
Relevant applications of transformation optics are electromagnetic cloaking and
field concentrators \cite{Rahma}.

Real time tuning of the metamaterial effective electromagnetic response is evidently an important achievement both for conceiving active radiation steering devices and for achieving a dynamical regime where the considered structure can be adapted to a specific task.

Metamaterial tuning methods may be divided into three fundamental classes: (i)
circuit tuning methods such as the inclusions of varactor diodes to control the
resonant frequency of unit cells, (ii) geometrical tuning methods, which allow
to modify the electromagnetic response by changing the geometry (size,
orientation, period, etc) of the unit cell, (iii) material tuning methods where
a constituent material modifies its electromagnetic response under a suitable
external stimulus like light, electric and magnetic field (see \cite{Turpin}
and references therein). Clearly, any tuning technique is suitable for a
specific frequency range.

In the Terahertz (THz) frequency region, some researchers are focusing their
attention on metamaterial devices containing semiconductor inclusions. Here the
device tunability stems from the fact that the semiconductor dielectric
response can be adjusted through different mechanisms such as photocarrier
injection \cite{Padilla_1,Shen_1,Rizza_2}, application of a bias voltage
\cite{Chen_1} and thermal excitation \cite{Han}. This allows to conceive a
novel class of photo-generated metamaterials realized without microfabrication
processes.

In the case of photocarrier injection, an optical beam with carrier frequency
into the absorbtion band generates electron-hole pairs. By spatially modulating
the optical beam one can generate a carrier grating able to produce the desired
value of permittivity. In \cite{Rizza_3} it was shown that a subwavelength
photo-generated grating in a GaAs slab can induce a THz metamaterial response
ranging from birefringent to hyperbolic to anisotropic negative dielectric.
Photo-induced metamaterials have also been used to manipulate the polarization
of THz pulses with subcycle switch-on times \cite{Kamaraju}, while active
control of THz optical activity has been achieved by means of chiral
photo-induced metamaterials \cite{Kanda}.

Photo-induced metamaterials are characterized by a pronounced spatial and
temporal reconfigurability being obtained without any material structuring (as
opposed to the standard metamaterials requiring fabrication).

In this paper we show that the method of \cite{Rizza_3} can be extended to the
microwave regime where, to the best of our knowledge, photoinduced
metamaterials have not yet been studied, although it is worth noting that
several authors proposed millimeter-wave manipulation by light control of a
semiconductor substrate \cite{Lee_1,Manasson_1,Manasson_2,Gallacher}. We stress
that reconfigurable photo-induced metamaterials in the microwave regime can
offer significant potential to achieve novel electromagnetic multifunction
applications such as, for example, dynamically photo-excited low reflection
metamaterials and shielding panels \cite{DePaulis}, adaptive perfect lenses
\cite{lens}, active polarizers \cite{Kamaraju} and actively controllable
cloaking devices \cite{cloak}.

We consider a photo-induced grating, imprinted into a silicon slab by using a
spatial light modulator (SLM), whose period is much smaller than the gigahertz
(GHz) wavelength. We analytically describe the coupling of optical beam and
free carriers and we derive the effective permittivity tensor describing the
homogenized metamaterial response. We specifically focus on the impact of
semiconductor parameters (diffusion $D$, free carrier life-time $\tau$, surface
recombination velocity $S$) on the homogenized response. Finally, thanks to the
vast tunability offered by the setup, we show that the proposed photoinduced
metamaterials can exhibit an extremely anisotropic response.

The proposed approach to obtain and test reconfigurable metamaterials consists
of a two-stage process: (i) an optical writing phase and (ii) a microwave
readout phase. As depicted in Fig. 1, an optical beam is modulated through a
SLM and it is launched onto a silicon slab. (i) The optical beam generates free
charges and \emph{writes} a carrier photo-induced grating. (ii) In a second
stage, the gigahertz waves \emph{reads} the imprinted dielectric modulation. By
changing the optical illumination, one can reconfigure period and depth of the
induced grating to obtain the desired effective microwave response.

The paper is organized as follows. In Sec. II, we describe the spatial dynamics of light and the generated photo-carriers (writing phase). In Sec. III, we
discuss the gigahertz response of the photo-induced metamaterials (readout phase). In Sec. IV we consider the impact of the semiconductor parameters on
the effective medium response. In Sec. V we compare the predictions of the considered metamaterial description with the results of full-wave simulations.
In Sec. VI we draw the conclusions.

\section{Writing phase}

Let us consider a transverse monochromatic optical field (TE) ${\bf E}=\mathrm{Re}[E(x,z)e^{i(k_0z-\omega t)}] \hat {\bf e}_y$ at an angular frequency $\omega$ which, after passing through a spatial light modulator (SLM), impinges on a silicon slab (Si) whose refractive index is $n=n_b+i\alpha/(2 k_0)$, where $n_b$ is the background refractive index, $\alpha$ is the absorbtion coefficient, and $k_0=\omega/c$. In our description of the scattering process of the optical TE wave launched onto the semiconductor, we assume that carrier dynamics do not affect light propagation and the diffraction length is much greater than both the slab thickness and the distance of the spatial light modulator to the vacuum-semiconductor interface.
\begin{figure}
\begin{center}
\includegraphics[width=0.5\textwidth]{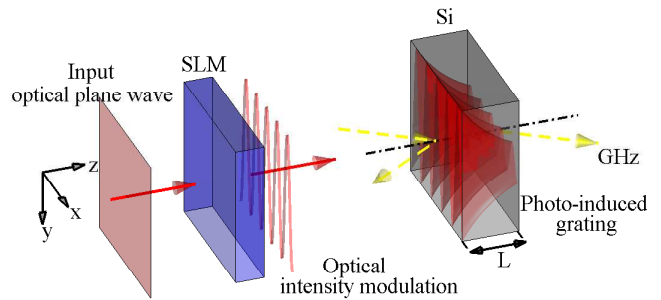}
\caption{(Color online) Sketch of a reconfigurable photoinduced metamaterial
and waves scattering geometry. An optical plane wave is modulated by a spatial
light modulator (SLM) and induces a dielectric grating into a silicon slab (Si)
of thickness $L$. Gigahertz (GHz) waves impinge onto the silicon slab (dashed
arrows).} \label{Fig1}
\end{center}
\end{figure}

Under these assumptions, in Appendix \ref{appA} we evaluate the light intensity
distribution within the semiconductor slab $I_\mathrm{slab}$ produced by an
arbitrary periodic 1D intensity profile $I_\mathrm{in}$ of the impinging
optical radiation. The optical beam excites electron-hole pairs and in Appendix
\ref{appB} we evaluate the photo-induced carrier density $\Delta p$ (see Eqs.
(\ref{trial}) and (\ref{trial2})) from the ambipolar equation (\ref{rate_eq})
for the intensity profile $I_\mathrm{slab}$ in the situation where the
dependence of both ambipolar diffusion parameter $D$ and carrier life-time
$\tau$ on carriers densities can be neglected. The appropriate boundary
conditions for the ambipolar equation are $D\,\partial_z \Delta p(x,z=0)=S\,
\Delta p(x,z=0)$, $D\,\partial_z \Delta p(x,z=L)=-S\,\Delta p(x,z=L)$ where $S$
is the surface recombination velocity. Therefore, in our scheme we do not
neglect carrier recombination due to impurity centers at the slab surface. As
we will show below, such surface effects have a significant impact on the
metamaterial effective response.

We focus on the specific and relevant impinging optical illumination
\begin{equation}\label{int}
I_\mathrm{in}=I^{(0)} \left[1+\cos\left(K_g x \right) \right]\,,
\end{equation}
where $I^{(0)}$ is the average intensity, $K_g=2 \pi/L_g$ and $L_g$ is the modulation period produced by the SLM.
The expressions obtained in the Appendices are valid for any slab thickness. However, in the following we will assume that
the semiconductor slab thickness is much smaller than the skin depth $1/\alpha$, i.e. $\alpha L\ll 1$.
In this limit we can set equal to 1 the exponential terms in Eq. (\ref{slow_2}) and, by inserting Eq. (\ref{int}) in that equation
we obtain the following expression for the optical intensity inside the silicon slab
\begin{equation}\label{I_slab}
I_\mathrm{slab}= n_b I^{(0)} \left(|t|^2+|r|^2 \right) \left[1+\cos\left( K_g x\right) \right]\,.
\end{equation}
The reflection and transmission coefficients $t$ and $r$ are complex constants reported in Eqs. (\ref{slow_1}).
In the limit $\alpha L\ll1$ one can further assume that the carrier density profile is uniform along the $z$ direction so that the photo-induced
grating according to Eq. (\ref{trial}) is
\begin{equation}\label{cariche}
\Delta p=\Delta p^{(0)}+2 \Delta p^{(1)} \cos\left( K_g x\right)\,,
\end{equation}
with
\begin{eqnarray}\label{dp_0_1}
\Delta p^{(0)}&=&\frac{Z_0\epsilon_0 n_b\alpha I^{(0)}} {\hbar k_0 D} \frac{q_2^{(0)}-|t|^2-|r|^2}{\alpha^2-{L_0}^{-2}}\,, \\
\Delta p^{(1)}&=&\frac{Z_0\epsilon_0 n_b\alpha I^{(0)}}{2\hbar k_0 D}
\frac{q_2^{(1)}-|t|^2-|r|^2}{\alpha^2-{L_1}^{-2}}\,.
\end{eqnarray}
The expressions of the effective diffusion lengths $L_m$ and of the constants $q_2^{(0)}$ and $q_2^{(1)}$ are given by Eqs. (\ref{ldm}) and  (\ref{q2}), respectively, in Appendix \ref{appB} ($Z_0$ is the vacuum impedance, $\hbar$ is the reduced Plank's constant).
\section{Readout phase}
In the readout phase we consider a monochromatic plane wave GHz field (at an
angular frequency $\Omega$) normally impinging onto the silicon slab interface,
namely ${\bf E}^\mathrm{(GHz)}=\mathrm{Re} \left[{\bf E}_0^\mathrm{(GHz)}
{\mathrm e}^{i(K_0 z-\Omega t)} \right]$ where ${\bf E}_0^\mathrm{(GHz)}$ is a
vector lying in the $xy$ plane and $K_0=\Omega/c$. The GHz dielectric
permittivity is described by the Drude model
\begin{equation}\label{GHz_die}
\epsilon^\mathrm{(GHz)}=\epsilon_{\infty}
-\frac{e^2}{\epsilon_0\Omega}\left[\frac{n}{m_n(\Omega+i
\gamma_n)}+\frac{p}{m_p(\Omega+i \gamma_p)}\right]\,,
\end{equation}
where $\epsilon_{\infty}$ is the background dielectric constant, $n$ $(p)$ is the electron (hole) density, $m_n$ $(m_p)$ is the associated effective mass, $\gamma_n$ $(\gamma_p)$ is the inverse of the relaxation time $\tau_n$ $(\tau_p)$, $e$ is the electric charge unit.
\begin{figure}
\begin{center}

\includegraphics[width=0.5\textwidth]{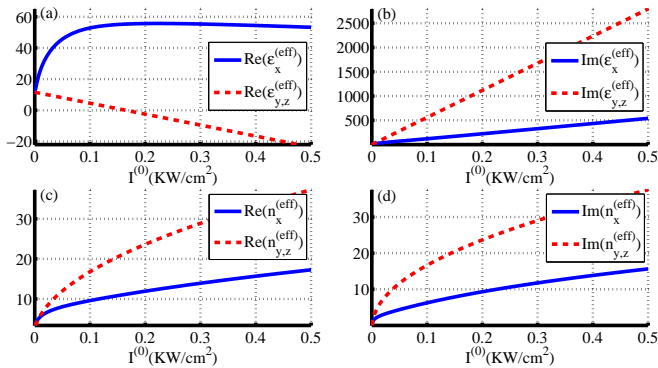}
\caption{(Color online) Real and imaginary part of the effective dielectric
permittivities $\epsilon_x^\mathrm{(eff)}$ and $\epsilon_{y,z}^\mathrm{(eff)}$
(a,b) and of the refractive indices $n_x^\mathrm{(eff)}$ and
$n_{y,z}^\mathrm{(eff)}$ (c,d) as functions of the optical intensity $I^{(0)}$,
with $L=1$ $\mu$m, $S=1$ m/s, $\tau=10^{-3}$ s.}
\end{center}
\end{figure}
Here we consider the homogenized regime where the period $L_g$ of the photoinduced grating is smaller than the microwave wavelength
$\lambda^\mathrm{(GHz)}=2 \pi c/ \Omega$. In this regime, two GHz plane waves linearly polarized parallel and orthogonal to the $x$ axis generally experience different dielectric responses. Through a multiscale technique \cite{Felbacq}, the effective dielectric permittivity tensor
$\epsilon^\mathrm{(eff)}=\mathrm{diag}[\epsilon_x^\mathrm{(eff)},\,\epsilon_{y,z}^\mathrm{(eff)},\,\epsilon_{y,z}^\mathrm{(eff)}]$ is obtained, where
\begin{equation}\label{homo}
\epsilon_x^\mathrm{(eff)}     = \left \langle \frac{1}{\epsilon^\mathrm{(GHz)}} \right \rangle^{-1}, \quad
\epsilon_{y,z}^\mathrm{(eff)} = \left \langle \epsilon^\mathrm{(GHz)} \right \rangle
\end{equation}
and $\langle f \rangle$ is the average of $f$ along the $x$-axis. Considering
that $p=p_0+\Delta p$, $n=n_0+\Delta n \simeq n_0+\Delta p$ ($p_0$ and $n_0$
are the hole and electron background contributions, respectively) and
substituting Eq. (\ref{cariche}) into Eq. (\ref{GHz_die}) we obtain
\begin{equation}\label{ep_GHz}
\epsilon^\mathrm{(GHz)}(x) = \epsilon^{(0)} + 2\epsilon^{(1)}\cos(K_g x)\,,
\end{equation}
with
\begin{eqnarray} \label{delr}
\epsilon^{(0)} &=& \epsilon_{\infty}-\frac{e^2}{\epsilon_0\Omega}
   \left[  \frac{1}{m_n}\frac{n_0+\Delta p^{(0)}}{\Omega+i\gamma_n} +\frac{1}{m_p}\frac{p_0+\Delta p^{(0)}}{\Omega+i\gamma_p} \right]\,, \nonumber \\
\epsilon^{(1)} &=& -\frac{e^2}{\epsilon_0
\Omega}\left[\frac{1}{m_n}\frac{\Delta p^{(1)}}{\Omega+i \gamma_n}
+\frac{1}{m_p}\frac{\Delta p^{(1)}}{\Omega+i \gamma_p} \right]\,.
\end{eqnarray}
Finally, by substituting Eq. (\ref{ep_GHz}) into Eq. (\ref{homo}) and performing the spatial averages, we obtain the effective dielectric permittivities components
\begin{equation} \label{effective}
\epsilon_x^\mathrm{(eff)} = \pm \sqrt{[\epsilon^{(0)}]^2-4
[\epsilon^{(1)}]^2}\,,\quad \epsilon_{y,z}^\mathrm{(eff)} = \epsilon^{(0)}\,.
\end{equation}
where the sign in the first equation is chosen in such a way that the imaginary
part of the effective permittivity is positive as for any passive media. We
hereafter set $\omega=2.90 \cdot 10^{15}$ rad/s, $\Omega=6.28 \cdot 10^{10}$
rad/s (corresponding to a frequency of $10$ GHz and a wavelength
$\lambda^\mathrm{(GHz)}=0.03$ m),
$L_g=\lambda^\mathrm{(GHz)}/20=1.5\cdot10^{-3}$ m, and we consider the
following parameters for the silicon slab: $n_b=3.8491$, $\alpha=3.0565 \cdot
10^{4}$ m$^{-1}$ \cite{Vuye}, $D=0.0022$ m$^2$/s, $n_0=10^{18}$ m$^{-3}$,
$p_0=0$ m$^{-3}$, $m_n=0.27\,m_0$, $m_p=0.37\,m_0$ ($m_0$ is the electron mass)
\cite{Riffe}, $\tau_n=2.3 \cdot 10^{-13}$ s, $\tau_p=1.3 \cdot 10^{-13}$ s,
$\epsilon_\infty=11.6$.

In Fig. 2 we report the effective parameters characterizing the electromagnetic
response as functions of the optical pump intensity with $L=1$ $\mu$m, $S=1$
m/s, $\tau=10^{-3}$ s. More precisely, in Figs. 2(a,b) we plot the effective
dielectric permittivities $\epsilon_{x}^\mathrm{(eff)}$,
$\epsilon_{y,z}^\mathrm{(eff)}$, and in Figs. 2(c,d) the refractive indices
$n_x^\mathrm{(eff)}$ and $n_{y,z}^\mathrm{(eff)}$.

Note that in Fig. 2(a) the real part of the dielectric permittivity
$\epsilon_{x}^\mathrm{(eff)}$ reaches its maximum value
$\epsilon_{x}^\mathrm{(eff)}=55.7$ at $I^{(0)}=0.22$ kW/cm$^2$ whereas the real
part of the dielectric permittivity $\epsilon_{y,z}^\mathrm{(eff)}$ changes
sign at $I^{(0)} \simeq 0.2$ kW/cm$^2$ so that the metamaterial has a
hyperbolic response for $I^{(0)} > 0.2$ kW/cm$^2$ .

The marked dependence on the average input intensity $I^{(0)}$ of the
dielectric permittivity coefficients displayed in Fig. 2 is a very important
feature supporting the reconfigurability of the proposed metamaterial setup. In
addition, the difference between $\epsilon_{x}^\mathrm{(eff)}$ and
$\epsilon_{y,z}^\mathrm{(eff)}$ significantly increases with $I^{(0)}$,
providing the effective medium response a huge uniaxial anisotropy.
\section{Dependence of the effective medium response on semiconductor parameters}
When $D \simeq 0$, carrier recombination has to compensate photo-generation and the photo-generated carrier density is given by
\begin{equation}
\Delta p=\tau \frac{\epsilon_0 \alpha Z_0}{\hbar k_0} I_\mathrm{slab}\,,
\end{equation}
(see Appendix \ref{appB}) namely $\Delta p$ is proportional to $\tau$. Hence,
the free carrier life-time becomes a fundamental parameter for achieving an
efficient coupling between light and free-carriers.

In Fig. 3 we plot the effective dielectric permittivities (with $L=1$ $\mu$m,
$S=1$ m/s) for three different values of that parameter. For $\tau=10^{-6}$ s
(solid line) the real parts of the effective permittivities are positive in the
considered intensity range since the photo-generated carrier modulation $\Delta
p$ is not sufficiently deep to produce metallic regions. For increasing values
of the free carrier life-time ($\tau=10^{-4}$ s (dashed line), $\tau=10^{-2}$ s
(dash-dot line)) the effective medium response anisotropy (with possible
hyperbolicity) accordingly grows.
\begin{figure}
\begin{center}
\includegraphics[width=0.5\textwidth]{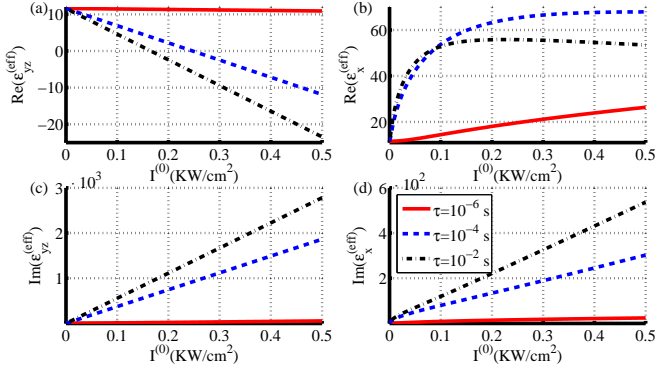}
\caption{(Color online) $\epsilon_{y,z}^\mathrm{(eff)}$,
$\epsilon_x^\mathrm{(eff)}$ as functions of the optical intensity $I^{(0)}$
with $L=1$ $\mu$m, $S=1$ m/s, for different recombination times: $\tau=10^{-6}$
s (solid line), $\tau=10^{-4}$ s (dashed line), $\tau=10^{-2}$ s (dash-dot
line).}
\end{center}
\end{figure}
\begin{figure}
\begin{center}
\includegraphics[width=0.5\textwidth]{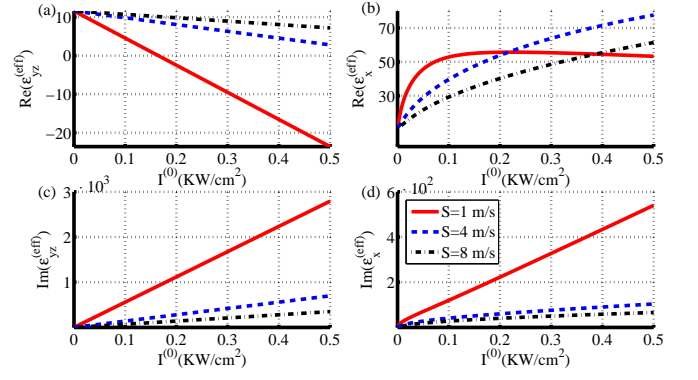}
\caption{(Color online) $\epsilon_{y,z}^\mathrm{(eff)}$,
$\epsilon_x^\mathrm{(eff)}$ as functions of the optical intensity $I^{(0)}$
with $L=1$ $\mu$m, $\tau=10^{-3}$ s, for different surface recombination
velocities: $S=1$ m/s (solid line), $S=4$ m/s (dashed line), $S=8$ m/s
(dash-dot line).}
\end{center}
\end{figure}
\begin{figure}
\begin{center}
\includegraphics[width=0.5\textwidth]{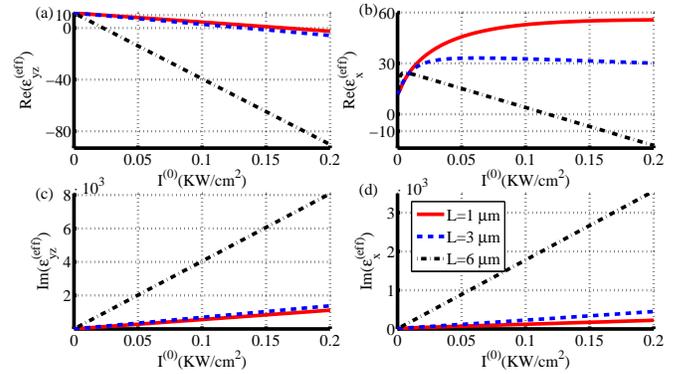}
\caption{(Color online) $\epsilon_{y,z}^\mathrm{(eff)}$,
$\epsilon_x^\mathrm{(eff)}$ as functions of the optical intensity $I^{(0)}$
with $S=1$ m/s, $\tau=10^{-3}$ s for different sample lengths: $L=1$ $\mu$m
(solid line), $L=3$ $\mu$m (dashed line), $L=6$ $\mu$m (dash-dot line).}
\end{center}
\end{figure}

Furthermore the settlement of the free-carrier modulation $\Delta p$ is
affected by the surface recombination velocity $S$ and by the semiconductor
thickness $L$. The presence of surface impurities reduces the density of the
photo-induced carrier and can not be generally neglected.

In Fig. 4 we plot the effective dielectric permittivities as functions of the average intensity, for $S=1$ m/s (solid line), $S=4$ m/s (dashed line), $S=8$ m/s (dash-dot line). For low recombination velocities ($S=1$ m/s), the optical pump illumination produces a deep carrier modulation and the metamaterial anisotropy is accordingly large to the point of showing hyperbolicity (for $I_0 > 0.16$ KW/cm$^2$). On the other hand, for higher recombination velocities ($S=4$ m/s, $S=8$ m/s), the photo-induced grating is shallow and the real part of the effective permittivity $\epsilon_{y,z}^\mathrm{(eff)}$ does not change sign in the considered range of illumination intensity ($0<I^{(0)}< 0.5$ kW/cm$^2$).

Finally, in Fig. 5 we plot the effective dielectric permittivities as functions
of the slab thickness $L$. Such a dependence is specific of the semiconductor
based metamaterial we are considering in this paper and it arises from the
recombination process occurring at the slab edges. The contribution of the
surface recombination is more pronounced the thinner is the semiconductor
sample, therefore thinner samples show smaller effective dielectric anisotropy.

\section{Full-wave simulations}
The predictions of the theoretical description illustrated in the previous
sections have been checked by evaluating numerically the GHz transmissivity of
a semiconductor slab hosting a photo-induced grating. The 2D full-wave
simulations have been performed with the comsol RF module \cite{com}. The
integration domain is composed by a semiconductor slab sandwiched by two vacuum
layers introduced to provide external excitation and to evaluate the
transmissivity.

We have solved Maxwell's equations for the GHz field (with the Drude dielectric
permittivity of  Eq. (\ref{GHz_die})) coupled to the ambipolar diffusion
equation (\ref{rate_eq}) with constant $D$ and $\tau$. At the entrance and exit
faces (orthogonal to the $z$ axis) we used matched boundary conditions on the
GHz field to excite a plane wave normally impinging onto the semiconductor slab
whereas we imposed the mixed boundary conditions $D\,\partial_z \Delta
p(x,z=0)=S\,\Delta p(x,z=0)$, $D\,\partial_z \Delta p(x,z=L)=-S\,\Delta
p(x,z=L)$ at the vacuum-semiconductor interfaces. Only one period along the
$x$-direction was considered by imposing the continuity conditions at the sides
of the unit cell. The numerical simulations have been performed for the
parameters used in Fig. 2.

In Fig. 6 we report the slab transmissivities obtained from numerical
simulation (circles and squares) for impinging waves linearly polarized along
the $x$-axis ($T_x$) and along $y$-axis ($T_{y,z}$) and compare them with the
transmissivity evaluated using the effective medium theory (solid and dashed
lines). In Fig. 6(a) we report such a comparison for the GHz wavelength
$\lambda^\mathrm{(GHz)}=0.3$ m and we note that numerical and analytical
results are in full agreement. In Fig. 6(b) the same comparison is drawn for
$\lambda^\mathrm{(GHz)}=0.03$ m and it reveals slight discrepancies between the
two predictions. We explain this discrepancy by observing that, in the first
case, the period-wavelength ratio is $L_g/\lambda^\mathrm{(GHz)}=200$ and the
medium dielectric response is fully homogenized whereas, in the second case,
$L_g/\lambda^\mathrm{(GHz)}=20$ so that dielectric homogenization is only
partially achieved and contributions due to spatial nonlocality play a role
\cite{Elser_1}.
\begin{figure}
\begin{center}
\includegraphics[width=0.5\textwidth]{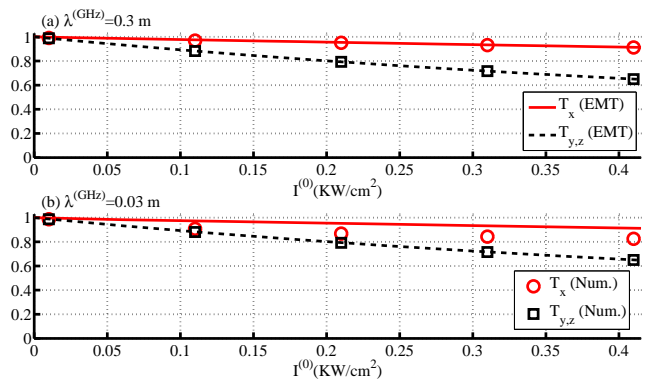}
\caption{(Color online) Comparison between the numerical transmissivity
(circles and squares) and those predicted by the effective medium theory (solid
and dashed lines) with $S=1$ m/s, $\tau=10^{-3}$ s, $L=1$ $\mu$m. We plot the
transmissivities $T_x$ and $T_{y,z}$ as functions of $I^{(0)}$ at
$\lambda^\mathrm{(GHz)}=0.3$ m (a) and at $\lambda^\mathrm{(GHz)}=0.03$ m (b).}
\end{center}
\end{figure}
\section{Conclusions}
In conclusions we have theoretically studied the GHz dielectric response of a metamaterial structure suitably photo-induced, and hence highly reconfigurable,
by an optical beam within a silicon slab. We have obtained analytical expressions for the components of the effective dielectric tensor in the
relevant case of sinusoidal light intensity modulation. We have shown that the GHz effective response has a marked dependence on the optical intensity and it
is generally extremely anisotropic. We have also investigated the impact of the semiconductor physical properties (free carrier life-time, surface
recombination velocity and slab thickness) on the metamaterial GHz effective response. In view of the high structural reconfigurability of the proposed system and of its extreme metamaterial anisotropy, we believe that our results can be useful to conceive a novel class of active devices to control the GHz field such as polarizers, polarization switchers, filters, lenses and spatial modulators.
\appendix
\section{Optical intensity distribution in the semiconductor slab} \label{appA}
As sketched in Fig. 1, we consider an optical plane wave spatially modulated along the $x$-axis by the SLM and normally impinging onto the vacuum-semiconductor interface (at the plane $z=0$), namely
\begin{equation}
\label{inc} {\bf E} ^{(i)}=F^{(i)}(x,z) e^{i \left(k_0 z- \omega t \right)} \hat {\bf e}_y.
\end{equation}
We here focus on the situation where the semiconductor slab thickness $L$ and the distance between the SLM and the vacuum-semiconductor interface $z=0$ are smaller than the impinging wave diffraction length $L_\mathrm{diffr} \approx k_0 n_b w^2/2$ (where $k_0=\omega/c$, $n_b$ is the real part of the refractive index and $w$ is the scale of the transverse modulation spatially imprinted by the SLM) so that the paraxial approach can be exploited and diffraction can be neglected.

The optical beam inside the semiconductor slab produces an inhomogeneous carrier density distribution (see Appendix \ref{appB}) which in turn affects optical propagation. Under the hypothesis of small refractive index modification, in our approach we neglect such a coupling. The field in the semiconductor slab can be represented as
\begin{equation}
\label{slow} {\bf E} = \left[F(x) e^{i k_0 n z }+ B(x) e^{- i k_0 n z }\right] e^{-i \omega t } \hat {\bf e}_y\,,
\end{equation}
where $F$ and $B$ are the slowly varying envelopes of the forward and the backward beams, $n=n_b+i \alpha/(2 k_0)$ is the semiconductor refractive index ($\alpha$ is the absorbtion coefficient). Imposing the continuity of electric and magnetic fields at the vacuum-semiconductor interfaces and neglecting diffraction, after some straightforward calculations we analytically obtain the two counterpropagating beams as functions of the incident field
$F(x)=t F^{(i)}(x,z=0)$ and $B(x)=r F^{(i)}(x,z=0)$ where
\begin{eqnarray}\label{slow_1}
t&=&\frac{2 (n+1) e^{-i k_0 n L}}{(n+1)^2e^{-i k_0 n L}-(n-1)^2e^{i k_0 n L}}\,,\nonumber \\
r&=&\frac{2 (n-1) e^{i k_0 n L}}{(n+1)^2e^{-i k_0 n L}-(n-1)^2e^{i k_0 n L}}\,,
\end{eqnarray}
so that the optical intensity distribution within the semiconductor slab $I_\mathrm{slab} = \frac{n_b}{2 Z_0} |{\bf E}|^2$ is
\begin{eqnarray}\label{slow_2}
I_\mathrm{slab} = n_b \left(|t|^2 e^{-\alpha z}+|r|^2 e^{\alpha z} \right) I_\mathrm{in}
\end{eqnarray}
where $I_\mathrm{in} =\frac{1}{2Z_0} |F^{(i)}(x,z=0)|^2$.
\section{Photo-generated carrier density distribution}\label{appB}
In the presence of electron-hole generation and recombination processes, the
photo-generated carrier density distribution satisfies the $2$-D continuity
equation or ambipolar equation \cite{solid_book}
\begin{equation}\label{rate_eq}
\frac{\partial}{\partial x} \left(D \frac{\partial \Delta p}{\partial x}
\right) +\frac{\partial}{\partial z} \left(D \frac{\partial \Delta p}{\partial
z} \right) + g_{eh}-r_{eh}=0,
\end{equation}
where $\Delta p$ is the excess hole density, $D=(n+p)/(n/D_p+p/D_n)$ is the
ambipolar diffusion coefficient ($D_i$ with $i=n,p$ are the diffusion
associated to the electron $n$ and hole $p$ densities), $g_{eh}$ and $r_{eh}$
are the electron-hole generation and recombination rates, respectively. Note
that the rate equation (\ref{rate_eq}) is obtained in the quasi-neutrality
condition, namely the excess electrons ($\Delta n$) and holes ($\Delta p$) are
quasi-equal and the total current density is near zero \cite{solid_book}. The
electron and hole densities are $n=n_0+\Delta n$, $p=p_0+\Delta p$,
respectively where $n_0$ and $p_0$ are the homogeneous background electron and
hole densities. The total generation-recombination rate is given by an external
contribution due to light absorbtion and a second one accounting for
recombination, i.e.
\begin{equation}
\label{g-r} g_{eh}-r_{eh} =  \frac{\epsilon_0 \alpha Z_0}{\hbar k_0}
I_\mathrm{slab} - \frac{\Delta p}{\tau}
\end{equation}
where $1/\tau=a+b \Delta p +c \Delta p^2$ and $a$ is the nonradiative
recombination rate coefficient, while the terms containing the coefficients $b$
and $c$ describe radiative transitions and Auger recombination, respectively
\cite{Garmire_1}.

We set $1/\tau = a$, assuming that $\Delta p$ is so small that nonlinear
contributions can be neglected, and $D=2/(1/D_n+1/D_p)$ assuming $n \simeq p$
(i.e. $\Delta n \simeq \Delta p$ and $n_0 \ll \Delta p$, $p_0 \ll \Delta p$).
Under these assumption Eq. (\ref{rate_eq}) becomes a linear equation which,
after using Eq. (\ref{g-r}) together with the intensity distribution of Eq.
(\ref{slow_2}), can be analytically solved. The periodicity of the optical beam
along the $x$-axis allows for a Fourier series expansion of the input intensity
distribution $I_\mathrm{in} = \frac{1}{2Z_0} \sum_{m=-\infty}^{+\infty} f^{(m)}
e^{i m K_g x}$ where $f^{(m)}=[f^{(-m)}]^*$. If we write the carrier
distribution as
\begin{equation}\label{trial}
\Delta p=\sum_{m=-\infty}^{+\infty} \Delta p^{(m)}(z) e^{i m K_g x},
\end{equation}
we get
\begin{eqnarray}\label{trial2}
\Delta p^{(m)}&=&\frac{\epsilon_0 n_b \alpha f^{(m)} }{2 \hbar k_0 D\left(\alpha^2-{L_m}^{-2}\right)}\left[ q_1^{(m)}\sinh\left(z/L_m\right) \right.  \\
&+& \left. q_2^{(m)}\cosh\left(z/L_m\right)- \left( |t|^2e^{-\alpha z}+|r|^2e^{\alpha z} \right) \right]\,, \nonumber
\end{eqnarray}
where the quantities
\begin{equation}\label{ldm}
L_m=L_0/\sqrt{1+\left(mK_gL_0\right)^2}\,,\quad L_0=\sqrt{D\tau}\,
\end{equation}
are effective diffusion lengths. The constant $q_1^{(m)}$, $q_2^{(m)}$ are
fixed by the boundary conditions. By imposing the mixed boundary conditions
$D\,\partial_z \Delta p(x,z=0)=S\,\Delta p(x,z=0)$ and $D\,\partial_z \Delta
p(x,z=L)=-S\,\Delta p(x,z=L)$ at the vacuum-semiconductor interfaces ($S$ is
the surface recombination velocity) we obtain
\begin{eqnarray}
q_1^{(m)}&=& \frac{1}{w}  \left\{\left[ S   \left(S-v_\alpha\right)e^{-\alpha L}- v_{1,m} \left(S+v_\alpha\right)\right] |t|^2 \right. \nonumber \\
&+& \left.  \left[ S   \left(S+v_\alpha\right) e^{\alpha L}- v_{1,m}\left(S-v_\alpha\right)\right] |r|^2 \right\}   \,,\label{q1} \\
q_2^{(m)}&=& \frac{1}{w}  \left\{\left[v_m\left(S-v_\alpha\right)e^{-\alpha L}+ v_{2,m} \left(S+v_\alpha\right)\right] |t|^2 \right. \nonumber \\
&+& \left.  \left[v_m\left(S+v_\alpha\right) e^{\alpha L}+ v_{2,m}\left(S-v_\alpha\right)\right] |r|^2 \right\}\,, \label{q2}
\end{eqnarray}
where $v_\alpha=D\alpha$, $v_m=D/L_m$, and
\begin{eqnarray}
v_{1,m} &=& v_m\sinh\left(L/L_m\right)+S\cosh\left(L/L_m\right)\,,\\
v_{2,m} &=& v_m\cosh\left(L/L_m\right)+S\sinh \left(L/L_m\right)\,,\\
w       &=& \left(v_m^2+S^2\right)\sinh\left(L/L_m\right) \nonumber\\
        &&  + 2v_mS \cosh\left(L/L_m\right)\,.
\end{eqnarray}

\end{document}